# Monolayer sorption of neon in mesoporous silica glass as monitored by WAXS

Duncan Kilburn, Paul E Sokol

Indiana University Cyclotron Facility, 2401 Milo B. Sampson Lane, Bloomington,

Indiana, 47404, USA

**Abstract**

We report measurements of the X-ray scattering intensity as mesoporous silica glasses are filled with Neon. The intensity of the first peak in the liquid-like diffraction pattern increases non-linearly with mass adsorbed. We outline a simple model assuming that the major coherent contribution to the first peak in the scattering function S($Q$) is due to interference from nearest-neighbour scatterers. This allows us to demonstrate an approach for surface area determination which does not rely on thermodynamic models – and is therefore complimentary to existing methods. We also suggest that the over-estimation of surface area by the traditional BET method may be resolved by using the capillary, and not the bulk, condensation pressure as the reference pressure, $p_0$. Furthermore, the alternative analysis offers an insight into the atomic structure of monatomic sorption, which may be of use for further studies on materials with different surface properties.

# I. INTRODUCTION

Mesoporous silica glasses are ideal systems for the study of matter in restricted geometries. Typically the silica encloses a network of open pores with diameters in the range 10-100 nm and specific surface areas which can be on the order of 1000 m$^2$/g. The classical method for characterizing such glasses involves gas sorption experiments which are analyzed according to thermodynamic models. The pertinent phenomenon for determining pore diameters is typically capillary condensation; that is, the suppression of a gas's saturation vapour pressure due to a concave curvature of the condensed phase. Historically, this was expressed in the Kelvin equation and then developed to account for adsorption-desorption hysteresis in the BJH [1] method for pore size distribution determination. More recently, modified versions of the Kelvin equation have been shown to accurately describe the pore diameters in ordered porous glass [2]. The classical determination of the surface area of a porous glass rests on analyzing the initial sorption and using models of the equilibrium between adsorbed layers to calculate the number of moles of a gas it takes to cover the surface. The BET method generalized Langmuir adsorption to multilayers and is still the default way to calculate surface areas [3].

There have been many studies on confined phases using Vycor porous glass which have reported numerous findings, notably suppression of phase transition temperatures [4] and altered phase structure in confinement [5]. Recently, a study has shown that in the hysteresis sorption region the mobility of an adsorbate is of a fundamentally different nature from outside that region [6]. The method for making Vycor (via spinodal decomposition) results in irregularly shaped pores which nevertheless have a relatively narrow size distribution. Using the method of molecular templating, it is possible to create porous glasses whose pores are ordered. For example, MCM 41 glass has cylindrical pores in a regular hexagonal structure. SBA 15 is similar but has additional open channels connecting the main pores. As these pores are arranged in a regular lattice, small angle X-ray scattering (SAXS) can be used to accurately measure the repeat distance of the pores [7,8,9,10,11]. As an extension of this, SAXS scans have been carried out simultaneously with a pore filling isotherm. The variation in the intensity of the peaks due to the long-range hexagonal pore lattice is modeled to extract information about the structure of the adsorbate filling those pores [12,13]. A layering structure similar to those invoked in thermodynamic models was reported. Similar experiments have also shown a slight expansion of the glass matrix structure upon nitrogen sorption [14]. In addition, wide angle X-ray scattering experiments have measured the atomic structure of a condensate within a mesoporous structure. These have focused mostly on phase transitions within pores, see for example Ref [5]; although some have monitored the full diffraction pattern as adsorption and then capillary condensation occurs [15,16]. For example, Huber and Knorr report diffraction patterns for adsorption of Argon in 75 Å pores at 65 K, which is below the confined freezing point [15]. They find that below bi-layer coverage, the diffraction pattern suggests that the layers form a triangular lattice with coherence length of the order of 16 Å. Above this, the bulk crystalline diffraction peak positions are reached at a coverage of about three monolayers. In addition, theoretical treatments of the thermodynamics of capillary condensation and equilibrium continue to generate much interest [17,18].

In this study, we report on data showing wide angle X-ray diffraction intensities as adsorption progresses for gas in two ordered mesoporous glasses. The gas adsorbed is Neon and is therefore a simple monatomic system and is comparable to the studies mentioned above (Ar and Kr). All isotherms are recorded at temperatures above the pore melting temperature so the sorption in all cases is leading to a liquid structure when capillary condensation occurs.

## II. EXPERIMENTAL DETAILS

Porous silica glasses were prepared by Prof. S. Komarneni of Pennsylvania State University. Sample A was manufactured following the method established for MCM 41 glasses: a reactive gel with a composition of 1.0 $SiO_2$:0.48 CTAB: 0.39 $Na_2O$: 0.29 $H_2SO_4$: 73.7 $H_2O$ was prepared by dispersing 14 g sodium silicate solution (~27% $SiO_2$) in 75 g of distilled water containing 1.76 g of $H_2SO_4$ and 10.8 g of CTAB under vigorous stirring. The obtained gel was then subjected to microwave-hydrothermal treatment at $100^oC$ for 5 hours in order to obtain highly ordered MCM 41 phase. The phase purity and crystallinity of the obtained samples were established with powder x-ray diffraction, nitrogen adsorption-desorption measurements and TEM analysis techniques. The method for preparing sample B is described in Ref. [19], it follows a non-standard method to achieve SBA-15 type glass.

Adsorption isotherms were measured using standard volumetric techniques. MKS Baratron differential pressure gauges allowed precision determination to ± 0.1 torr corresponding to an accuracy of 1 % in sample filling. Prior to each experiment, the sample was evacuated using a turbo pump for ~12 hours at 350 K.

The X-ray diffraction patterns were obtained with an RU-200 X-ray generator using a Silver rotating anode source operated at 45 kV and 70 mA. A standard θ-2θ scattering geometry was employed. A graphite monochromater on the scattered beam was used to select the unresolved Ag $K_\alpha$ with wavelength 0.561 Å. The 22 keV X-ray beam could easily penetrate through the sample and experiments were performed in transmission geometry using a sample cell with a thickness parallel to the incident beam of 1.52 mm. Beryllium windows on the sample contributed crystalline diffraction peaks to the recorded pattern which were removed prior to analysis. The diffraction intensities were also corrected for absorption by the sample and polarization effects [20].

### III.     RESULTS AND DISCUSSION

Sorption isotherms for the two systems investigated are shown in Fig. 1. They are both similar to isotherms measured in equivalent systems previously. Both glasses exhibit type IV isotherms according to BET classification [21] which is characteristic of mesoporous materials. The differences between the two, however, are quite clear and are indicative of properties of the two systems. Most noticeably, the region due to capillary condensation covers a distinctly smaller relative pressure range, and is at a lower relative pressure, for sample A; these two observations are due to a narrower pore size distribution and lower mean pore size, respectively. It is likely that there is a contribution of secondary mesopores to sample B's isotherm. The total specific surface areas calculated using the standard BET method [3] are 1060 m$^2$/g (A) and 630 m$^2$/g (B), the corresponding glass filling (mass) fractions are listed in table I. The area covered by a Neon atom was calculated using a hexagonal area per atom (*i.e.* assuming solid-like structure for monolayer sorption) with nearest-neighbour distance 3.13 Å [20]. The hole pore radii distributions were calculated using a modified Kelvin equation method [10] with the standard Nitrogen isotherms (not shown). As mentioned previously, sample A has a very narrow pore size distribution with a mean diameter of 42 Å, sample B has a larger distribution of hole sizes: it is roughly a top-hat function extending from a diameter of 80 to 200 Å, resulting in a mean of 140 Å.

The total pore volume is calculated from the total volume of neon in the pores after capillary condensation. This assumes that the density of the neon is not affected by confinement, which is confirmed by comparing structure factors for bulk and confined neon. The volumes are 1.17 cm$^3$/g and 0.77 cm$^3$/g for samples A and B, respectively.

The scattering pattern is shown in figure 2 for empty and full sample A; the peaks are due to the beryllium windows. The pattern for the empty mesoporous glass has a disordered structure at higher $Q$ values whilst the increase at low $Q$ signals the beginning of a peak due to the longer-range structure of the pores themselves, which are at comparable values to those reported previously for this class of glass [11]. For the glass whose pores are filled with neon the diffraction pattern at higher $Q$ values has a contribution from the liquid neon superimposed on top. It can be seen that the largest difference is at a scattering vector of 2.45 Å$^{-1}$, which corresponds to the first peak in the scattering function for liquid neon at 27 K [23]. The reduction in scattering intensity at small $Q$ for the neon-filled pores is due to a reduction in scattering cross-section contrast on the length scale of the pore repeat distance.

Measuring the adsorption isotherm in Fig 1 whilst simultaneously monitoring the intensity of the peak height at $Q = 2.45$ Å$^{-1}$ allows the intensity of that peak to be plotted as a function of the amount of gas adsorbed. This first peak in the scattering function is primarily influenced by interference from nearest neighbours scattering, and we focus hereafter on this peak. This is shown for both glasses in Fig. 3. The intensity is the total intensity measured from the sample as it is filled minus the initial intensity of the empty pore matrix. We assume that as this Q value is characteristic of correlations at the neon-neon nearest neighbor distance, it is valid to subtract the background in this way since the host matrix electron correlations at this length-scale will be unaffected by the additional neon. It is important to note that this is the intensity of the peak due to neon only and so the points tend towards the origin. For simplicity in this paper we now focus on monatomic liquids, but many of the arguments used are valid generally. The X-ray scattering intensity for bulk monatomic liquids at a given $Q$ is proportional to the number of atoms scattering [24]. The scattering function of a system of particles can be expressed as being the sum of two components: one originating due to scattering from independent particles; the other being due to interference effects from multiple particles. The scattering from independent particles incorporates an incoherent fraction, which results when the X-rays lose a small amount of energy on scattering, and an independent coherent fraction, which is expressed as the form factor. For the purposes of the present analysis the distinction between these contributions to the independent scattering component are irrelevant; it is enough to recognize that they contain no information about the relative positions of different atoms, unlike the multiple-particle scattering component. In the present system there will also be contributions to this scattered intensity due to correlations between electrons in the silica glass surface and the monolayer of neon atoms. This contribution, however, will increase linearly with the number of neon atoms at the Q-value specified and so can be considered as being part of the independent coherent fraction for the purposes of this paper. For a simple bulk monatomic liquid, the scattering intensity can be expressed as

$$I(Q) = Nk(\sigma_m S_m(Q) + \sigma_i S_i(Q)), \qquad (1)$$

where $I(Q)$ is the intensity measured, $N$ is the number of particles (in this case, Ne atoms), $\sigma$ is the scattering cross section and $S$ is the scattering function, where the sucbscripts 'm' and 'i' denote multiple-particle and independent scattering, respectively. $k$ is a constant which depends on the characteristics of the X-ray diffractometer settings. From Fig. 3 it can be seen that above glass filling (mass) fractions of $x=1.2$ (A) and $x=0.8$ (B) the intensity is proportional to the mass of Ne, demonstrating bulk-like scattering behaviour. Fits of these plots to $I(Q = 2.45) = Nk_{bulk}$ (for simplicity of notation, for the rest of the paper $k_{bulk} = k(\sigma_m S_m(Q = 2.45) + \sigma_i S_i(Q = 2.45)))$ gives $k_{bulk}$ listed in table I. The flattening of the curves at higher $x$ values is due to the condensation of 'true' bulk Ne in the sample cell which occurs initially at the bottom of the cell, out of the path of the X-ray beam, and hence does not contribute to an increase in $I$. The region below the bulk-like scattering corresponds to scattering from non-bulk neon, which clearly does not follow Eq. (1), but instead requires some generalized form. For sample A there is a larger difference between the extrapolated plot of $I(Q= 2.45) = Nk_{\text{bulk}}$ (shown as a dashed line) and the data than there is for sample B. These differences are due to the different pore geometry, size and surface structure for the two samples. It is well established that for mesoporous materials, the sorption of gas progresses initially via a monolayer, then a bilayer, and so on until capillary condensation occurs. In the rest of this paper we will focus on the initial part of this sorption process and describe how plots of this kind can give useful information about the adsorbing surface.

If the adsorption of atoms onto a surface occurs with a random spatial distribution then the initial atoms will be unlikely to have any adjacent neighbours in the distance range typified by nearest neighbours in bulk. The X-ray scattering due to interference from multiple atoms from such an array will be very low. The last atoms adsorbed into a monolayer, however, will all have nearest neighbours, next nearest neighbours and so on and so will all contribute the maximum possible to the intensity scattering from a monolayer. To correctly model the scattering intensity with monolayer adsorption, therefore, the scattering function describing interference effects from multiple atoms must be calculated as a function of $N$, the number of neon atoms:

$$I(Q) = Nk(\sigma_m S_m(Q,N) + \sigma_i S_i(Q)), \tag{2}$$

If the assumption is made that the dominant contribution to the scattering intensity due to interference from multiple atoms comes from nearest neighbour atoms then this can be done relatively simply. Firstly, we assign the value $\alpha$ to the contribution a nearest neighbour pair makes to $k\sigma_m S_m(Q,N)$. Next, the expected number of nearest neighbours for an adsorbed atom is expressed as:

$$\langle \text{n.n.} \rangle = y \langle P_{n,t}(y) \rangle = y \frac{\sum_{n \leq t} P_{n,t} n}{\sum_{n \leq t} P_{n,t}}, \tag{3}$$

where $P_{n,t}$ is the probability that a given site has $n$ nearest neighbours, and $y = N/N_{\text{mono}}$, $N_{\text{mono}}$ being the number of atoms required to fill the monolayer. To calculate $P_{n,t}$ we picture the adsorption as occurring on a surface where each of the adsorption sites mentioned above has on average $t$ nearest neighbours, which gives: $P_{n,t} = y^n(1-y)^{t-n}$. $\langle P_{n,t}(y) \rangle$ then is the expected number of occupied nearest neighbour sites for a given site (the reference site itself is not necessarily occupied) and varies between 0 and $t$. Figure 4 shows $\langle P_{n,t}(y) \rangle/t$, normalized to $\langle P_{n,t}(y=1) \rangle = t$, as a function of $y$. Initially, we assume the monolayer sorption takes place on a triangular lattice, similar to the (111) plane of a bulk fcc lattice (coordination number, $t = 6$) [25]. The total scattering function for a partial monolayer is then given by:

$$I(Q) = N[\langle P_{n,6}(y) \rangle \alpha + \beta]. \tag{4}$$

The contribution of the independent scattering that each atom makes to $I(Q)$ is assigned the value $N\beta$. In this simple model, the multiple-atom scattering function for bulk neon is given by the expression $k\sigma_m S_m(Q) = 12\alpha$, i.e. we assume 12 nearest neighbours. The independent scattering function contribution per atom for bulk is unchanged from its value in the monolayer, $\beta$. This latter point is likely to be an approximation: as noted above, there will be some contribution from silica-neon electron correlations to the independent scattering in the monolayer. We assume, however, that this contribution is small because it involves different atomic species and so does not contribute as strongly as the same species neon-neon correlations. An additional concern is the possible effect of a pore wall corona or microporosity in the sample. There has been a considerable amount of literature addressing this issue in porous silica materials, particularly SBA-15 [13,26,27]. Whilst this additional variation in structure on the length-scale of the order of the pore-repeat distance is of concern for small angle scattering measurements, we are only concerned with there being a definable surface between matrix and pore (whether macro- or micro-) that the neon atoms can adsorb onto. In terms of the relation for bulk neon, this means that $k_{bulk} = 12\alpha + \beta$. Figure 5 shows an enlarged section of the initial part of Fig. 3, where the intensity of the peak at 2.45 Å$^{-1}$ is given as a function of the mass of neon for the first monolayer region, that is below $x_{mono}$. For sample B, the intensities are shown from a more detailed scan of this region than the data in Fig.3, but the two scans show similar variation in this region; for sample A, a detailed scan was performed for the whole isotherm and so the data in Figs 3 and 5 are identical. The parameters $\alpha$, $\beta$ and $x_{mono}$ from non-linear least squares fits to Eq. 4 for samples A and B are shown in table I; the fits as shown in Fig. 5 have reduced $\chi^2$ values of 1.24 and 1.7, corresponding to probabilities 0.15 and 0.015, respectively. It can be seen from table I that the data is in reasonable agreement with the relation $k_{bulk} = 12\alpha + \beta$, bearing in mind that determination of $k_{bulk}$ and $12\alpha + \beta$ are entirely independent of each other. This last point must be considered carefully. From Fig. 4 it can be seen that the normalized curves of $<P_{n,t}(y)>$ are similar. Given the level of statistics from our experiments, we cannot reasonably differentiate between such distributions, assuming $t$ does not vary too far from 6 (for non-integer $t$ the sum in Eq. 3 is replaced with an integral). This means that the equality ($k_{bulk} - \beta$)/ $\alpha = 12$ simply fixes the ratio of nearest neighbours in bulk to those the monolayer at 12:6. If we were to make the assumption that there were 8.4 nearest neighbours in bulk liquid neon, as was calculated previously from X-ray measurements [23], then the ratio above would suggest 4.2 nearest neighbours in the monolayer. The values of $x_{mono}$ from the fit are shown in table I. The value for sample B is close to that from BET analysis whereas that for sample A is significantly less. It has been reported many times before that BET analysis over-estimates by approximately 10-15% the surface area of mesoporous materials whose diameters are in the tens of angstroms range [10,11,28-30]. This has been explained as being the result of several effects: firstly, surface heterogeneity [10] has been suggested as a reason, although why this should necessarily lead to an over-estimate is unclear; secondly, as the pore sizes decrease, the pressure range in which capillary condensation occurs impinges upon the pressure range in which the BET equation is fit (typically ~0.05 – 0.3 $p/p_0$) [10,11,28]; thirdly, there appears to be a overlap in the formation of the first and second layer of adsorbate[10]. In the original derivation of the BET equation [3] $p_0$ is the saturation vapour pressure of the gas or, equivalently, the pressure at which the number of adsorbed layers becomes infinite. We suggest that for the highly curved pore surfaces such as those found in mesoporous glasses this condition is not met (as is usually implicitly assumed) at the bulk condensation vapour pressure, but is instead met at the capillary condensation vapour pressure. As we are considering a pore a few nanometers in diameter the number of adsorbed layers is restricted spatially, but thermodynamically the condition of infinite layers can be considered as being met at the capillary condensation pressure, $p_c$. Re-calculating $x_{mono}$ using BET analysis with $p_c$ instead of $p_0$ as the limiting pressure gives values of 0.347 ± 0.001 and 0.239 ± 0.001 for samples A and B, respectively. The new value for A is now in agreement with that derived from the X-ray measurements (Table I), whilst that for B is still in reasonable agreement. For the sample B it is difficult to know what pressure to use for $p_c$, given the wide range covered by capillary condensation. We have used the pressure mid-way through the capillary condensation, but it may be that a pore-size weighted average should be used for more precise calculation. The approach to calculating BET surface areas for mesoporous materials described here is still likely to be an approximation, albeit slightly less crude than previous efforts. Indeed, the limitations listed previously (surface heterogeneity etc.) are still, in principle, valid. In the modified approach suggested here it is also probably correct to separate the surface area into fractions that are internal and external to the porous silica powder grains – it still seems reasonable to use $p_0$ as the limiting pressure for the external surface area under the BET method. In addition, the derivation of the BET equation assumes that the evaporation-condensation properties of the molecules in the second and higher adsorbed layers are equal and are the same as those of the liquid state. For angstrom-sized pores, the curvature of each adsorbed layer will be different leading to different evaporation-condensation properties – something which is not explicitly accounted for in the current approach.

One form of the BET equation is:

$$\frac{v}{v_{mono}} = \frac{c(p/p_0)}{(1-p/p_0)(1-p/p_0+c(p/p_0))} \tag{5}$$

Where $v/v_{mono}$ is the degree of surface coverage and $c$ is related to the difference between the heat of adsorption for the first layer and all subsequent layers. In principle, therefore, given a set of systems with identical surface chemistry and roughness adsorbing the same molecules, plots of $v/v_{mono}$ against $p/p_0$ should superimpose. Kruk *et al.* have plotted these values in Figure 6. of Ref.[11] for a series of nitrogen adsorption isotherms on MCM 41 glasses with a range of pore sizes and find that there is a systematic shifting of the plots along the $p/p_0$ axis. If, however, $p_c$ is used in the plot instead of $p_0$ (taking values from Table 2. of the same paper), the data do superimpose to within experimental error. This is taken as further evidence that it is the condensation vapour pressure of a gas in the vicinity of the surface in question, and not necessarily the bulk condensation vapour pressure, which should be used as the reference pressure for BET analysis.

## IV. CONCLUSIONS

We have presented X-ray scattering data collected as Neon is adsorbed onto the surface of porous silica glasses and then undergoes capillary condensation. A simple model was then developed which allows the surface area of the glasses to be determined. As this determination uses a structural rather than a thermodynamic model it is complimentary to existing, commonly used methods, such as BET surface area determination. Furthermore, we have used the new insight provided by this model to address the known problem of overestimation of surface areas by existing thermodynamic methods. This is a simple model for a monatomic gas. It gives insights into, but is not expected to describe fully, the sorption process. For more complicated systems, including surface roughness, Ising-type interactions may be required for full analysis.

## V. ACKNOWLEDGEMENTS

This work was supported under Grant No. DE-FG02-01ER45912. The porous glasses were prepared by B. Newalkar and S. Komarneni of the Materials Research Laboratory, Pennsylvania State University. We would like to acknowledge valuable discussions with M. Cole.

**References:**

1. E. P. Barrett, L. G. Joyner, and P. P. Halenda, J. Am. Chem. Soc. **73(1)**, 373 (1951).
2. M. Kruk, M. Jaroniec and A Sayari, Langmuir **13**, 6267 (**1997**).
3. S. Brunauer, P. H. Emmett, and E. Teller, J. Am. Chem. Soc. **60(2)**, 309 (1938).
4. E. Molz, A. P. Y. Wong, M. H. W. Chan, and J. R. Beamish, Phys. Rev. B **48(9)**, 5741 (1993).
5. D. W. Brown, P. E. Sokol and S. N. Ehrlich, Phys. Rev. Lett. **81(5)**, 1019 (1998).
6. R. Valiullin et.al. Nature **443**, 965 (2006).
7. C. T. Kresge, Nature **359**, 710 (1992).
8. J. S. Beck, J. Am. Chem. Soc.**114**, 10834 (1992).
9. K. Morishige, H. Fujii, M. Uga, and D. Kinukawa, Langmuir **13**, 3494 (1997).
10. M. Kruk, M. Jaroniec and A. Sayari, Langmuir **13**, 6267 (1997).
11. M. Kruk, M. Jaroniec and A. Sayari, J. Phys. Chem. B **101**, 583 (1997).
12. A. Ch. Mitropoulos, J. M. Haynes, R. M. Richardson, and N. K. Kanellopoulos, Phys. Rev. B **52(14)**, 10035 (1995).

13. G. A. Zickler, *et al.* Phys. Rev. B **73(18)**, 184109 (2006).

14. P-A. Albouy and A. Aryal, Chem. Mater. **14**, 3391 (2002).

15. P. Huber and K. Knorr, Phys. Rev. B **60(18)**, 12657 (1999).

16. K. Morishige, K. Kawano and T. Hayashigi, J. Phys. Chem. B **104**, 10298 (2000).

17. R. Evans, J. Phys.: Cond. Mat. **2**, 8989 (1990).

18. S. M. Gatica and M. W. Cole, Phys. Rev. E **72(4)**, 041602 (2005).

19. B. Newalkar, S. Komarneni and H. Katsuki, Chem. Comm. **23**, 2389 (2000).

20. H. P. Klug and L. E. Alexander, *X-ray Diffraction Procedures For Polycrystalline and Amorphous Materials*, 2nd ed, Chapter 12 (John Wiley and Sons: New York, 1974).

21. S. J. Gregg and K. S. W. Sing, *Adsorption, Surface Area and Porosity*, 1st ed (Academic Press: London, 1967).

22. D. G. Henshaw, Phys. Rev. **111(6)**, 1470 (1958).

23. D. Stripe and C. W. Tompson, J. Chem. Phys. **36(2)**, 392 (1962).

24. N. S. Gingrich, Rev. Mod. Phys. **15(1)**, 90 (1943).

25. P. Huber and K. Knorr, Phys. Rev. B **60(18)**, 12657 (1999).

26. M. Impéror-Clerc, P. Davidson, and A. Davidson, J. Am. Chem. Soc. **122**, 11925 (2000).

27. T. Hofmann, *et al.* Phys. Rev. B **72(6)**, 064122 (2005).

28. M. Kruk, M. Jaroniec and A. Sayari, Chem. Mater. **11**, 492 (1999).

29. M. Kruk, M. Jaroniec, R. Ryoo, and S-H Joo, Chem. Mater. **12**, 1414 (2000).

30. M. Thommes, R. Köhn, and M. Fröba, J. Phys. Chem. B **104**, 7932 (2000).

31. H. P. Klug and L. E. Alexander, *X-ray Diffraction Procedures For Polycrystalline and Amorphous Materials*, 2nd ed, Chapter 12 (John Wiley and Sons: New York, 1974).

**Tables**

**TABLE I. Fit parameters from adsorption isotherms. $x_{mono,BET}$ is the mass fraction of Neon for monolayer coverage from the BET equation, $x_{mono}$ is similar but from X-ray method. $\alpha$, $\beta$ and $k_{bulk}$ are fit parameters (to $I(Q = 2.45) = Nk_{bulk}$ and Eq. 4) from the monolayer and bulk sections of Fig. 3., respectively.**

| Sample | $x_{mono,BET}$ | $x_{mono}$ ($\pm 0.01$) | $\alpha$ ($\pm 10$) | $\beta$ ($\pm 50$) | $12\alpha + \beta$ ($\pm 130$) | $k_{bulk}$ |
|---|---|---|---|---|---|---|
| A | 0.419 | 0.35 | 217 | 2300 | 4910 | $5160 \pm 10$ |
| B | 0.249 | 0.26 | 273 | 3500 | 6776 | $6610 \pm 30$ |

**Captions to the figures.**

**FIG 1. Neon adsorption isotherms for samples A and B.**

**FIG 2. X-ray scattering pattern for sample A when empty and also when filled with liquid Neon. The sharp peaks are due to the Beryllium windows on the sample cell.**

**FIG 3. Plot of the intensity of the X-ray scattering peak at $Q = 2.45$ Å$^{-1}$ of Neon in the porous glasses. Arrows indicate limits of bulk condensation region.**

FIG 4. For monolayer adsorption on a lattice. $\langle P_{n,t}(y)\rangle$ is the expectation value of the number of occupied nearest neighbour sites where $t$ is the average number of nearest neighbour sites and $y$ is the fraction of the monolayer which is occupied.

FIG 5. Same as figure 3 but focused in on the monolayer region. Symbols are experimental data points. Full lines are fits to Eq. 4 with parameters given in table 1. Dashed lines are extrapolations of the initial rise of Intensity *vs*. *x* to emphasis the deviation from linearity as monolayer sorption continues.

**Figures**

Figure 1.

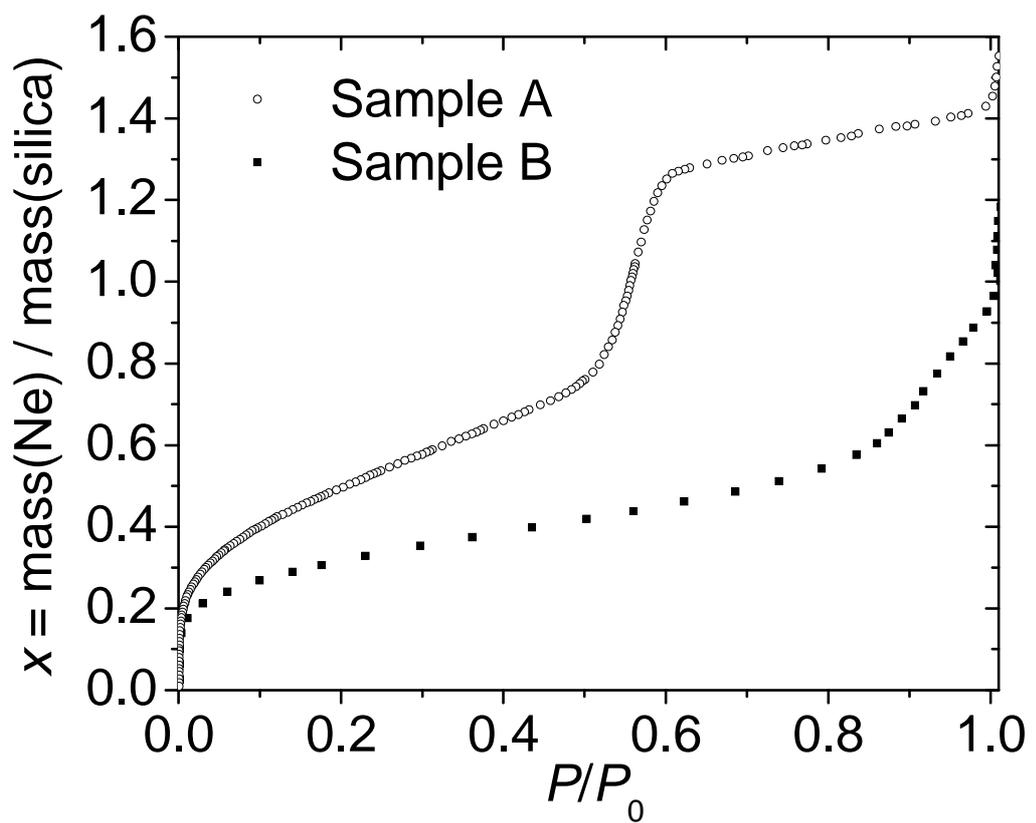

Figure 2

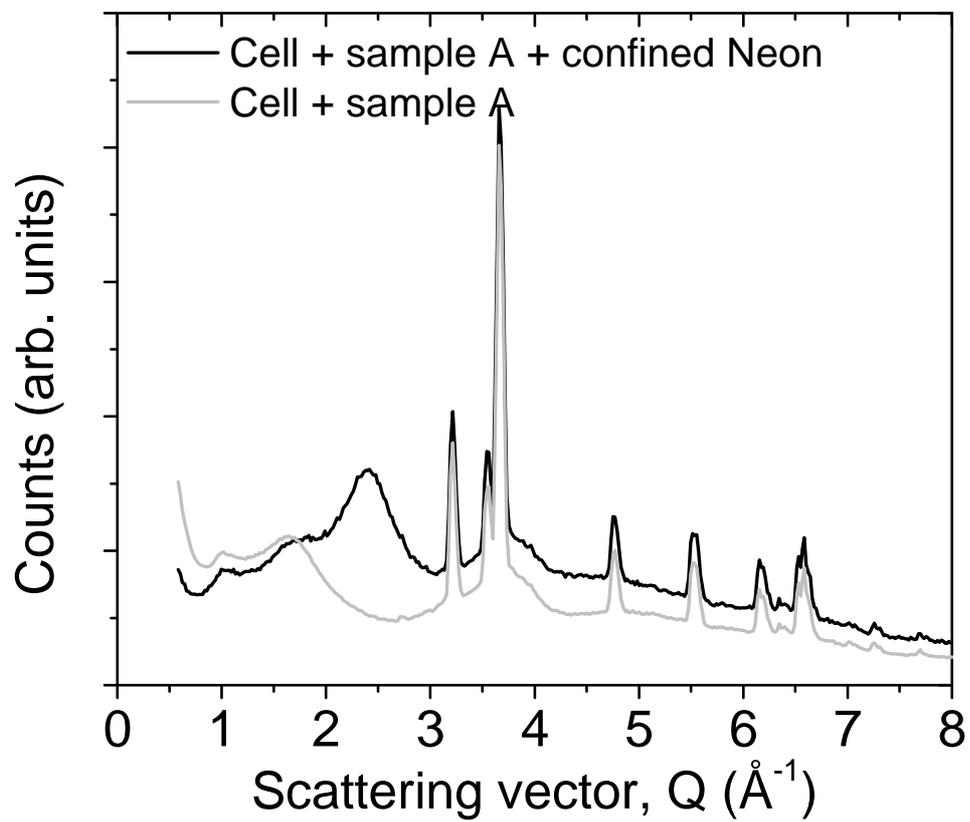

Figure 3.

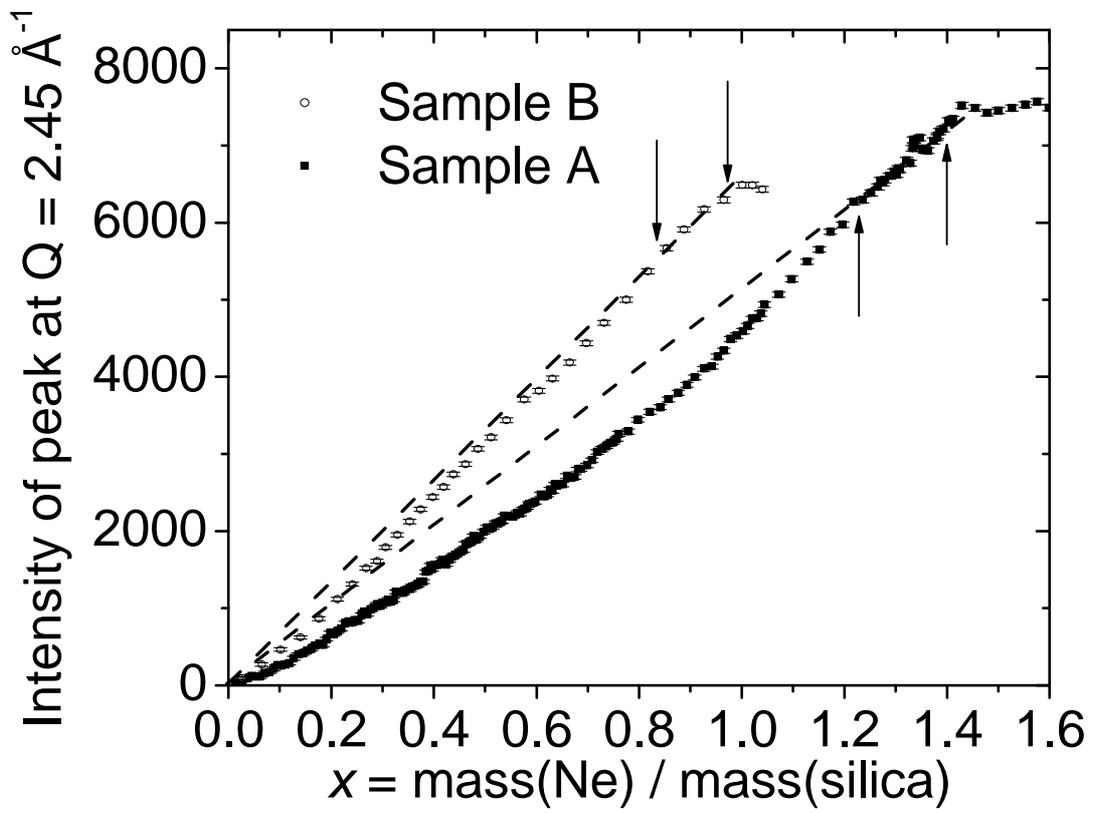

Figure4.

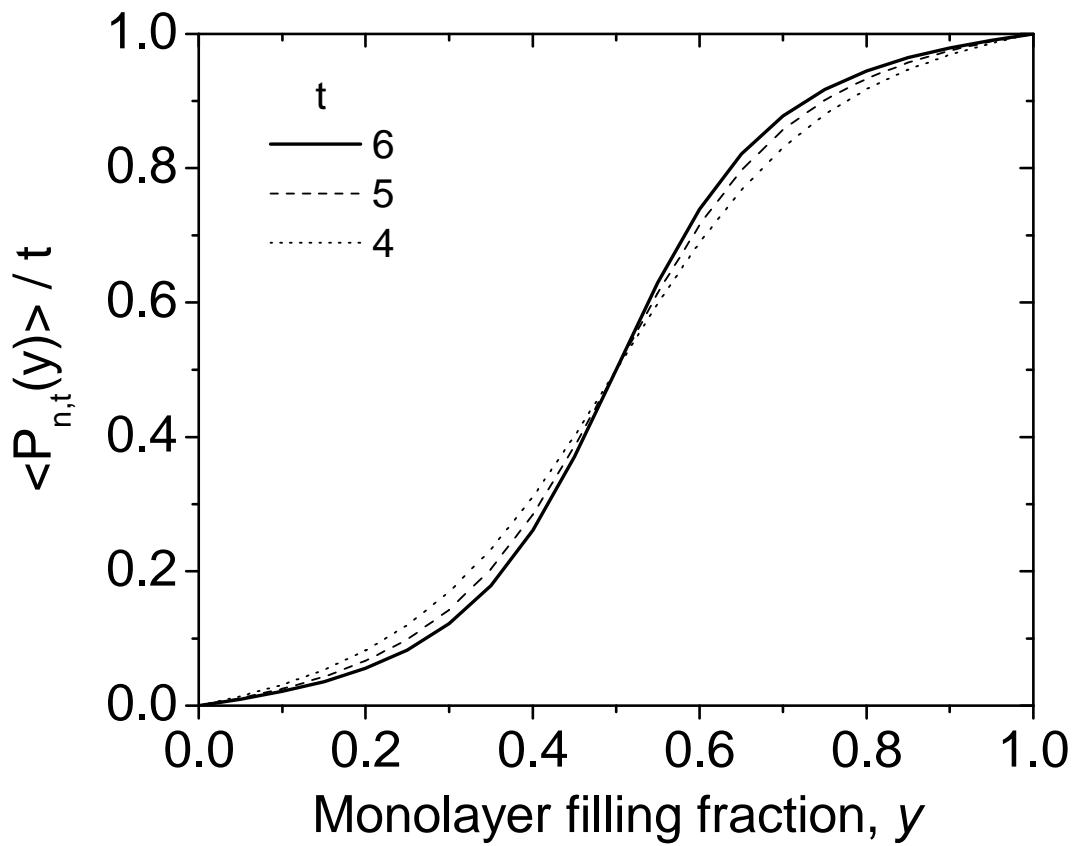

Figure 5.

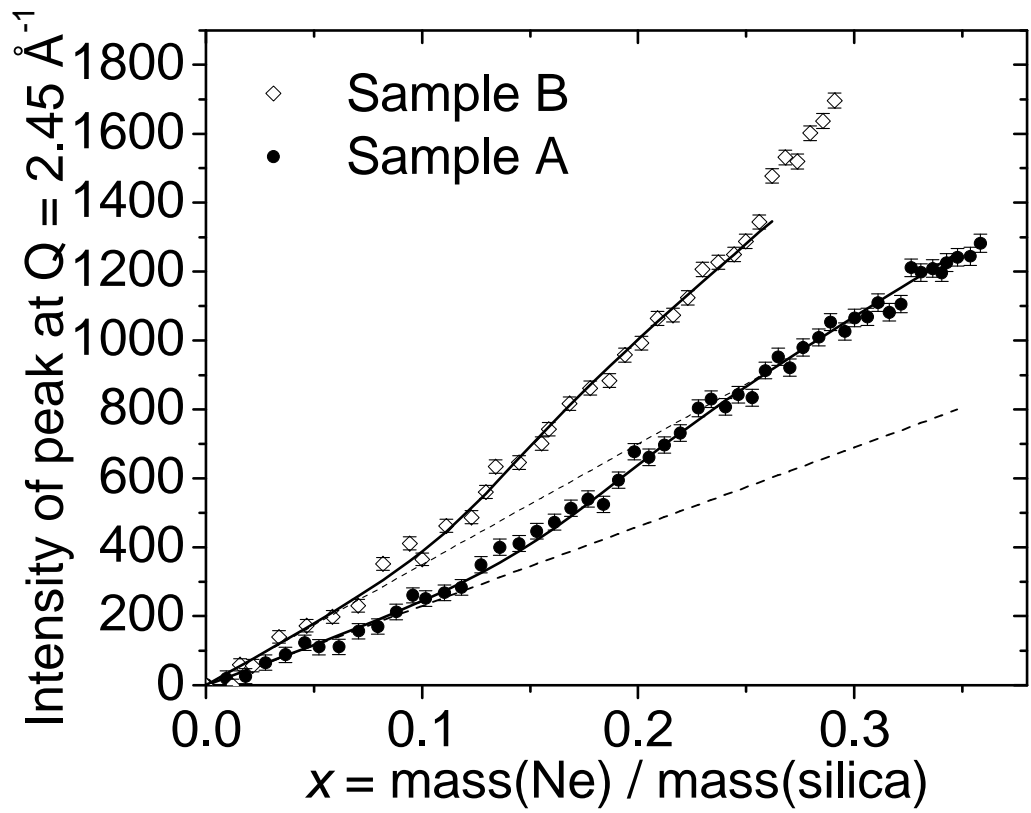